# On the Phase Transitions That Cannot Materialize

Yuri Mnyukh
76 Peggy lane, Farmington, CT 06032, USA, e-mail: *yuri@mnyukh.com*
(Dated: December 24, 2013)

___

**Abstract** The succession of suggested mechanisms of solid-state phase transitions − Second-order, Lambda, Martensitic, Displacive, Topological, Order-Disorder, Soft-mode, Incommensurate, Scaling and Quantum − are analyzed and explained why they cannot be realized in nature. All of them assume a *cooperative* structural rearrangement as opposed to the only real one which is simply a variant of the *crystal growth*. Like all kinds of crystal growth, a solid-state phase transition proceeds by *molecule-by-molecule* building the crystal of a different structure, while the surrounding original crystal is used as the building material.

___

## 1. Introduction

When contemplating possible mechanisms of solid state phase transitions, a care should be taken that they would not be inconsistent with thermodynamics. An infinitesimal change of the thermodynamic parameter ($dT$ in case of temperature) may produce only two results: either (A) an infinitesimal quantity of the new phase emerges, with the structure and properties changed by finite values, or (B) a physically infinitesimal "qualitative" change occurs uniformly throughout the whole macroscopic volume [1]. These conditions, however, are only necessary ones: they do not guarantee both versions to be found in nature.

## 2. Universal Crystal Growth *vs.* Second-Order Phase Transitions

There is no doubt that version 'A' is actually realized: it is an abstract description of the usually observed phase transitions by nucleation and growth. Every input of a minuscule quantity of heat $\delta Q$ either creates a nucleus or, if it exists, shifts the interface position by a minuscule length $\delta \ell$. The issue is, however, whether version 'B' can materialize. As far back as 1933, Ehrenfest classified phase transitions by *first-order* and *second-order*. The validity of the classification was disputed by Justi and Laue (the latter was a Noble Prize Laureate) who insisted that there is no thermodynamic or experimental justification for second-order phase transitions [2]. Landau [3,4], in disregard to those objections, developed a theory of second-order phase transitions. Landau and Lifshitz in their book "Statistical Physics" [5] devoted a special chapter to them, claiming that they "may also exist". Since then, it became widely accepted that there are "discontinuous" *first-order* phase transitions, exhibiting "jumps" in their physical properties, as well as "continuous" *second-order* phase transitions, showing no such jumps.

The properties of the second-order phase transitions were clearly stated. Such a transition occurs at a fixed *critical* (or *Curie*) point $T_c$ where the two crystal structures are identical. There they change *continuously*; only the crystal symmetry experiences a "jump". Neither overcooling nor overheating are possible (no *hysteresis*), nor liberation or absorption of heat can take place (no *latent heat*). Coincidence of the structure orientations goes without saying. These characteristics will help in the analysis of the phase transitions that do not materialize (Sections 4 - 12). In practice, all "second-order" phase transitions fail to fit them exactly.

Prior to considering the solid-state phase transitions that do not materialize, those which do materialize should be described. They were classified as *first order* and called "usual" by Landau. He defined them as a process when the *crystal structure changes abruptly, latent heat is absorbed or released, symmetries of the phases are not related, and overheating or overcooling is possible.* In his times their molecular mechanism was not discovered yet. Later on, the systematical experimental studies by this author and associates [6-19] revealed their physical nature. The transitions were fount to be a variant of *crystal growth*, very much analogous to crystal growth from liquids or gases, but this time from a crystal medium. The results were summarized in the book [20] and articles [21-24]. Specifics of the crystal growth in a crystal medium (after a peculiar "non-classical" nucleation) is illustrated by Fig. 1.

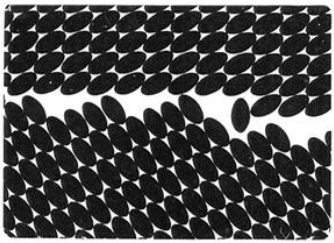

**Figure 1**. Molecular model of phase transition in a crystal. The *contact* interface is a rational crystal plane in the resultant phase, but not necessarily in the initial phase. The interface advancement has the *edgewise* mechanism. It proceeds by shuttle-like strokes of small steps (kinks), filled by molecule-by-molecule, and then layer-by-layer in this manner. The gap of 0.5 molecular layer (on average) is wide enough to provide steric freedom for the molecular relocation at the kink, but is sufficiently narrow for the relocation to occur under attraction from the side of resultant crystal.

The nucleation is heterogeneous, located in *optimum microcavities*. The activation temperature $T_n$ of each potential nucleus is encoded by the microcavity size and shape. All those temperatures are different and lagging relative to the temperature point $T_o$ where the free energies of the phases are equal. *Hysteresis $\Delta T_n = T_n - T_o$ is inevitable*, and not mere possible.

An essential result of the studies was the conclusion that second-order phase transitions do not exist. All prominent examples of "second order" phase transitions turned out to be erroneous. *Justi and Laue were right when contending that there is no thermodynamic or experimental justification for second-order phase transitions.*

The remaining non-reclassified "second-order" phase transitions were usually attributed to layered crystals. Phase transitions in layer crystals have been proven [16] to materialize by nucleation and growth, but its specific morphology made it easy to assign them second-order. A layered structure consists of strongly bounded, energetically advantageous two-dimensional units − molecular layers − appearing in both phases. There the interlayer interaction is weak on definition. Since the layer stacking contributes relatively little to the total lattice energy, the difference in the total free energies of the two structural variants is small, and so is the latent heat. Change from one polymorph to the other is reduced mainly to the mode of layer stacking. The layer parameters themselves are only slightly affected by the different layer stacking. In practice, layered structures always have numerous defects of imprecise layer stacking. Most of these defects are minute wedge-like interlayer cracks located at the crystal faces as viewed from the side of layer edges. In such a microcavity there always is a point where the gap has the optimum width for nucleation. There the molecular relocation from one wall to the other occurs with no steric hindrance and, at the same time, with the aid of attraction from the opposite wall. In view of the close structural similarity of the layers in the two polymorphs, *the nucleation is epitaxial* with a very small hysteresis. Orienting effect of the substrate (the opposite wall) preserves the orientation of molecular layers.

Now we can compare the characteristics of the *epitaxial* phase transitions with those of second-order phase transitions:

|  | *Second-order* | *Epitaxial* |
|---|---|---|
| ■ Structure orientations: | No change | layers: Same |
| ■ Structural similarity: | Identical | Very similar |
| ■ Latent heat: | Zero | Very small |
| ■ Hysteresis: | Zero | Very small |
| ■ Latent heat | Zero | Very small |

Epitaxial transition in DL-*norleucine* (DL-N) at ~117.2 °C [16] is an instructive example (see Fig. 2).

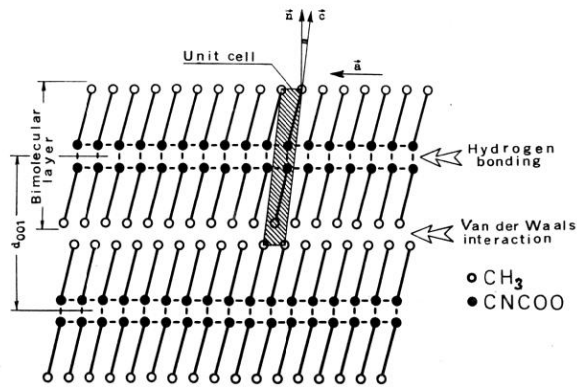

**Figure 2**. Characteristic features of the DL-*norleucine* (DL-N) crystal structure.

DL-N is a short-chain aliphatic substance
$$CH_3 \cdot (CH_2)_3 \cdot CHNH_2 \cdot COOH$$
with a layered crystal structure typical of chain molecules, where the molecular axes are quite or almost perpendicular to the layer plane. Each layer is bimolecular: the *CNCOO* groups of the molecules are pointed toward the center of the layer where they form a network of hydrogen bonds *N-H...O*. This central "skeleton" turns the bimolecular layer into a firm structural unit. The interlayer interaction is much weaker, because it is of a purely Van der Waals' type, so the layer stacking is governed exclusively by the principle of close packing. As a result, both DL-N polymorphs have a pronounced layered structure of almost the same layers in different stacking mode.

Without taking special precautions, it would be easy to assign it second order: it occurs "instantly", "without hysteresis" and change of crystal orientation. But careful experimental study [16] of its single crystals (they were thin lamellae parallel to the molecular layers) has revealed: (a) It materialized by moving interfaces over the lamellae; (b) Hysteresis $\Delta T_n$ was well detectable, but was only 0.2-0.8 $^{\circ}$C; (c) The orientation of the layers did not change; (c) Laue-patterns were almost identical; (d) The layer parameters remained almost same within 1%; (e) The quantitative ratio of the coexisting phases was changing from 0% to 100% over a small temperature range; (g) The long spacing (indicator of layer stacking) changed by 4.1%.

There is a single general molecular mechanism of all solid-state phase transitions: *nucleation and crystal growth*, formerly called "first order". It exhibits itself in two forms: *epitaxial* and *non-epitaxial*. It is the former that was erroneously taken for one or another "cooperative" mechanism of phase transitions.

## 3. 300 Mechanisms of One Phenomenon

It will take a long journey before the rearrangement shown in Fig.1 is accepted as the only real molecular mechanism of solid-state phase transitions. We were able to count in the literature more than 300 types/mechanisms of solid-state phase transitions. Even if they are sorted out into groups, their number does not lend credibility to all of them; rather it indicates the failure to identify the general one. Such a state of affairs is in keen contrast with what is known about nature's laws. Nature is thrifty. There is a single equilibrium state of any solid matter, be it a metal, ionic, or organic substance: it is a *crystal state*. Crystals can come into being from vapors, melts, solutions, or other crystals. There is only one general mechanism by which crystals of any nature can emerge from any solution, vapor, or melt: it is a *nucleation and growth*. This is hardly consistent with the idea that the same process in a solid medium requires scores of diverse mechanisms.

"Transition" means a *process*: passage from one state/condition to another. Giving a name to a phase transition means an identification of the specific *mechanism* of passage from one phase to another. This should be taken into account when looking at the collection of 300 different "mechanisms" listed in [20] (Appendix 1). Some of that chaos of names can be conditionally sorted out into groups. It is to be noted that the idea on a *cooperative* character of those mechanisms was always present, sometimes as open assumption, but mostly as a subconscious matter of course.

● Names somehow indicating at, rather than describing, the process (mechanism) of the phase transition: displacive, order-disorder, cooperative, diffusional, distortive, catastrophic, spin-flop, cation ordering, continuous… It is assumed that the phase transition is reduced to atomic/molecular displacements, structural distortion, spin-flopping, *etc*.

● Names having a more or less established theory of the mechanisms (however erroneous) in the literature: martensitic, soft mode, incommensurate, second order, quantum.

● Names carrying no characteristics at all, except being *not* something: "usual" are not martensitic, "classical" are not quantum, "structural" are not ferromagnetic, ferroelectric or superconducting, "diffusionless" are not diffusional… So are "ordinary", "normal" and "simple".

● Names of particular authors: Kastelein, Jahn-Teller, Mott, Anderson-Mott, Kosterlitz-Thouless, Berezinskii-Kosterlitz-Thouless, Ising, Lifshitz, Oguchi, Wilson, Stenley-Kaplan, Gardner, Neel, Peierls, Potts, Salam, Verway. This is a convenient way of identification: it is prestigious to those authors, absolves the responsibility to define them …and impedes scrutiny.

● Names of the driving forces, evidently in the belief that they identify specific phase transition mechanisms: density-driven, density-driven quantum, electronically driven, driven by soft-shear acoustic mode, driven by soft mode, current-induced, pressure-induced, shock-induced, stress-induced, field-induced. That belief is invalid, considering that phase transition is driven by imbalance of free energies, and the role of any driving force is only to affect the free energy.

● A loose group of names that are too formal to reflect meaningfully on the mechanism: first order (showing "jumps" in physical properties), lambda (showing singularity of the heat capacity reminiscent to letter 'lambda'), infinite order, weak-order, non-weak, isothermal, thermodynamic, non-thermodynamic, volume-change, symmetry-breaking, symmetric-antisymmetric.

● Names indicating the prominent property of the crystal: ferromagnetic, ferroelectric, superconducting.

The unifying idea that all that diversity is the effect of a single cause − changing of the crystal structure − was missing. The following sections concentrate on those of the suggested mechanisms that significantly affected science on phase transitions and are not still completely abandoned.

# 4. Lambda-Transitions

## 4.1. Everyone Believed It Is a Heat Capacity

The sharp peaks of heat capacity reminiscent to letter λ, recorded at the temperatures of solid-state phase transitions, challenged the theorists to explain their origin. The first λ-peak was observed by Simon in $NH_4Cl$ phase transition [25]. Later on, it was repeated many times and numerous other cases were reported. Thus, more than 30 experimental λ-peaks presented as "Specific heat $C_P$ of [substance] *vs.* temperature *T"* were shown in the book by Parsonage and Staveley [26]. The theories were unable to account for the phenomenon. P.W. Anderson wrote [27]: "Landau, just before his death, nominated [lambda-anomalies] as the most important as yet unsolved problem in theoretical physics, and many of us agreed with him… Experimental observations of singular behavior at critical points… multiplied as years went on… For instance, it have been observed that magnetization of ferromagnets and antiferromagnets appeared to vanish roughly as $(T_C-T)^{1/3}$ near the Curie point, and that the λ-point had a roughly logarithmitic specific heat $(T-T_C)^0$ nominally". Feynman stated [28] that "One of the challenges of theoretical physics today is to find an exact theoretical description of the character of the specific heat near the Curie transition - an intriguing problem which has not yet been solved."

This intriguing problem will be solved here. There were three main reasons for that theoretical impasse. (1) The λ-peaks were actually observed in first-, and not second-order phase transitions (including ferromagnetic transitions which are all "magnetostructural" [22]) (2) The first-order phase transitions exhibited latent heat, but it was mistaken for heat capacity. (3) An important limitation of the adiabatic calorimetry utilized in the measurements was unnoticed.

## 4.2. Reinterpretation of Old Experimental Data

The canonical case of "specific heat λ-anomaly" in $NH_4Cl$ around -30.6 °C will be re-examined. This case is of a special significance. It was the first where a λ-peak in specific heat measurements through a solid-state phase transition was reported and the only example used by Landau in his original articles on the theory of continuous second-order phase transitions [29]. This phase transition was a subject of numerous studies by different experimental techniques and considered most thoroughly investigated. In every calorimetric work (*e.g.*, [30-38]) a sharp λ-peak was recorded; neither author expressed doubts in a *specific heat* nature of the peak. The transition has been designated as a *cooperative order-disorder phase transition of the lambda type* and used to exemplify such a type of phase transitions. However, no one maintained that the λ-anomaly was understood.

It should be noted that many of the above-mentioned calorimetric studies were undertaken well after 1942 when the experimental work by Dinichert [39] was published. His work revealed that the transition in $NH_4Cl$ was spread over a temperature range where only mass fractions $m_L$ and $m_H$ of the two distinct L (low-temperature) and H (high-temperature) coexisting phases were changing, producing "sigmoid"-shaped curves. The direct and reverse runs formed a hysteresis loop Fig. 3(a). The fact that the phase transition is first-order was incontrovertible, but not identified as such.

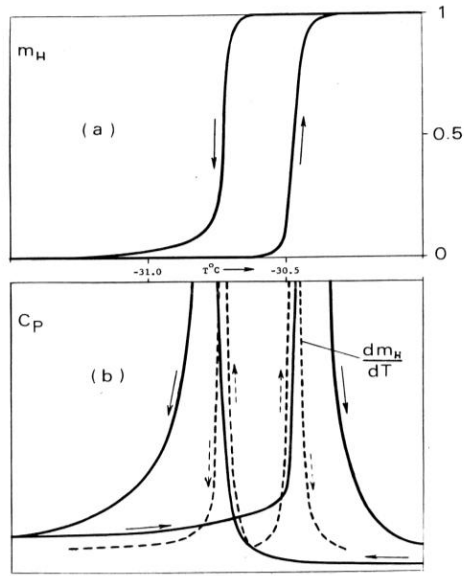

**Figure 3.** Phase transition in *NH₄Cl*.
(a) The hysteresis loop by Dinichert represents mass fraction of high-temperature phase, $m_H$, in the two-phase, *L+H*, range of transition; $m_L+m_H = 1$.
(b, solid lines) The λ-peaks from calorimetric measurements by Extermann and Weigle.
The plots are positioned under one another in the same temperature scale to make it evident that the shape of the peaks is proportional to *fist derivative* (dotted curves) of the $m_H(T)$.

In Fig. 3 the Dinichert's data are compared with the calorimetric measurements by Extermann and Weigle [32]. The latter exhibited "anomalies of heat capacity" (as the authors called the λ-peaks) and the hysteresis of the λ-peaks. Because of the hysteresis, it had already to become evident at this point (but was not) that the λ-peaks cannot be of a heat capacity, considering that heat capacity is a *unique function* of temperature. The graphs 'a' and 'b' are positioned under one another in the same temperature scale to reveal that the shape and location of the peaks are very close to *first derivative* of the $m_H(T)$ (dashed curves). It remains only to note that *latent heat* of the phase transition must be proportional to $dm_H/dT$. Thus, the latent heat of the first-order phase transition, lost in the numerous calorimetric studies, is found, eliminating the long-time theoretical mystery.

### 4.3. Limitations of Adiabatic Calorimetry

A legitimate question can be raised: why did not publication of the Dinichert's work change the λ-peaks interpretation from "heat capacity" to "latent heat"? The answer is: knowledge of the actual phase transition mechanism outlined in Section 2 was required. But there was also a secondary reason hidden in the calorimetric technique itself.

The goal of numerous calorimetric studies of λ-peaks in $NH_4Cl$ and other substances was to delineate shape of these peaks with the greatest possible precision. An adiabatic calorimetry, it seemed, suited best to achieve it. The adiabatic calorimeters, however, are only "one way" instruments in the sense the measurements can be carried out only as a function of increasing temperature. In the case under consideration, however, it was vital to perform both temperature-ascending and descending runs - otherwise existence of hysteresis would not be detected. And it was not detected. For example, in [37] the transition in $NH_4Cl$ was interpreted as occurring at the fixed temperature point $T_\lambda$ = 245.502 ± 0.004 K defined as a position of λ-peak. The high precision of measurements was useless: that $T_\lambda$ exceeded $T_o$ by 3°.

The results by Extermann and Weigle were not typical. The kind of calorimetry they utilized permitted both ascending and descending runs. That was a significant advantage over the adiabatic calorimetry used by others in the subsequent years. But there was also a shortcoming in their technique resulted in the unnoticed error in the presentation of the λ-peaks in Fig. 3b: the *exothermic latent heat* peak in the descending run had to be *negative* (looking downward).

### 4.4 Final Proof: It is Latent Heat

Differential scanning calorimetry (DSC) is free of the above shortcomings [40]. Carrying out temperature descending runs with DSC is as easy as ascending runs. Most importantly, it displays endothermic and exothermic peaks with *opposite* signs in the chart recordings, which results from the manner the signal is measured [20] (Appendix 2). If the λ-peak in $NH_4Cl$ is a *latent heat* of phase transition, as was concluded above, the peak in a descending run must be exothermic and look downward. Our strip-chart recordings made with a Perkin-Elmer DSC-1B instrument immediately revealed that the peak acquires opposite sign in the reverse run (Fig. 3). Its hysteresis was also unveiled.

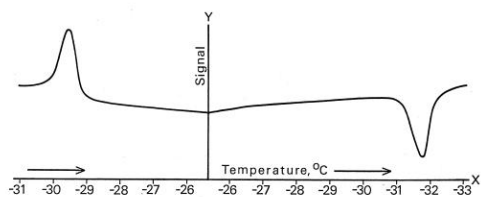

**Figure 4**. The actual DSC recording of $NH_4Cl$ phase transition cycle, displaying temperature-ascending and descending peaks as endothermic and exothermic accordingly, thus delivering final proof of a latent heat nature of the λ-peak [20] (Appendix 2)..

## 5. Martensitic Transformations

The "martensitic" mechanism of phase transitions was one of the oldest in the succession of the proposed different mechanisms. It came from physical metallurgists who studied formation of a phase called *martensit* in iron alloys from the higher-temperature phases. This mechanism was later claimed to cover many other solid-solid phase transitions. *Martensitic transformation* was assumed to be a strictly orderly process localized at a straight interface called "habit plane". There the two crystal structures exactly match with one another, the adjacent lattices on both sides of the habit plane being under local elastic distortions to provide this matching. A martensitic transformation occurs at a specific temperature $T_M$ which is neither $T_o$, nor $T_c$. The velocity of the interface propagation is that of a sound wave, rather than a function of temperature. There must be a certain rigorous orientation relationship (OR) between the crystal lattices prior to and after the transformation. The martensitic transformation, assumed to be a cooperative at interface, was theoretically approximated by a uniform transformation in the bulk. Since direct observation of phase transitions in iron and its alloys is an extremely difficult task, the suggested "martensitic" mechanism was based more on imagination than on solid facts.

The alternative to martensitic transformations was sometimes called *diffusional*, but diffusion was a too slow process to account for the rates of "non-martensitic" phase transitions. Then the terms "usual" or "nucleation and growth" were used. These terms were not descriptive at all. There was no room for second-order phase transitions in the classification.

The more "martensitic transformations" were investigated, the more it became evident that *they do not have a single specific experimental characteristic* separating them from what was claimed to be their alternative. They start from nucleation; their actual speed was lower than that of sound propagation and depended on temperature; temperature hysteresis was not their specific feature either; OR was not always as expected, or was not strict, or was absent. All attempts to find characteristics of specifically *martensitic* mechanism have failed. They very well matched to the *nucleation and growth* as presented in Section 2.

Once dominated over a significant part of literature, the *martensitic transformations*, as a specific phase transition mechanism, was fading for a period of time until it was recently somewhat resurrected in relation to the *shape memory* effect. Now it is taken for granted; the problems with its introduction and definition are forgotten. As to the shape memory, it is actually related to the *epitaxial* phase transitions [20] (Addendum F).

## 6. Displacive Phase Transitions

The *displacive* mechanism was put forward by Buerger [41,42] solely on the basis of comparison of the crystal structures before and after a phase transition. This author shared a common belief that it was sufficient to make judgment about its *process*. A rigorous OR was assumed, but not always verified.

Buerger suggested that structures can change into one another in two ways. If they are similar, the transition does not involve breakdown of the original bonding and is *displacive*. But, if there is no way to reform the initial crystal without breaking the existing bonding net, the transition must be *reconstructive*. The descriptions given to these two mechanisms were ambiguous. The *reconstructive* transitions are first-order, but actually assumed to be cooperative. "Their structures are so different that the only way a transformation can be effected is by disintegrating one structure into small units and constructing a new edifice from the units"; such transition is "sluggish", because the substance must pass through the intermediate state of a higher energy. It suffices to note that at that time there already were plenty of experimental data on phase transitions by propagation of interfaces, the fact not being taken into account.

The description of *displacive* phase transitions was not less problematic. They are fast, barrierless, involving only a small displacement of one or more kinds of the atoms. The problem was that most, if not all, cases were "hybrids" with some bonds had to be broken. We were informed that many *displacive* transitions exhibit a small energy jump, certainly indicating first-order phase transition, but the physical rearrangement could still proceed as in second-order phase transitions. Such "firstsecond"-order hybrid phase transitions are not allowed by thermodynamics (see Section 1).

There were more drawbacks. The introduction of the two distinct types − *displacive* and *reconstructive* − turned out to be only a headline for a rather cumbersome classification. It was found impossible to relate them with the changes in the first and second coordination in the structure. Several mechanisms, such as "dilatational" and "rotational", were added. They were neither quite *displacive*, nor quite *reconstructive*. Finally, the predicted velocities of phase transitions ("rapid" or "sluggish") did not correlate with experiment. (As McCrone [43] pointed out, "one should always be ready to meet unforeseen velocities"). The whole effort was a geometrical exercise. There was no attempts to invoke thermodynamics.

If *displacive* phase transitions could exist, the DL-N (Fig. 2) could be their best example. The OR was preserved. The resultant structure could be imagined as the initial one with its rigid molecular layers simply slipped to the new mode of layer stacking. It has been proven, however, that in order to produce the almost identical new molecular layers, every original one was a subject of full molecule-by-molecule reconstruction.

## 7. Topological Phase Transitions

*Topological* phase transitions are a sophisticated version of the *displacive* ones. There are phase transitions, plenty of them, which even most inventive theorists would unable to squeeze into the "displacive" category. The mechanism of these "reconstructive" first-order phase transitions cried for explanation. The topology, a branch of mathematics, was called for help.

The "topological" approach was based on the conviction that the resultant crystal must be a *modification* of the initial one. A *cooperative continuous* character of the process was a matter of course, so there was no need to look into the experimental literature. A possibility of molecule-by-molecule reconstruction to the crystallographically independent structure did not come to mind.

So, if not by simple displacement, than how? The answer was: phase transitions proceed through several topological stages of displacements / deformations / distortions. The geometry of the participating crystal structures is analyzed and if an imaginary pass can be suggested, it is declared to be the phase transition mechanism in that particular case. Then the efforts could turn to finding the individual phase transition mechanism in next case in the same manner.

## 8. Order-Disorder Phase Transitions

Phase transitions in which all or some constituent molecules, or their parts, of a crystal loose their definite orientations due to thermal agitation are called *order-disorder*. The resultant state was given name *orientation-disordered crystals* (ODCs). Some authors divide phase transitions into two broad types: *order-disorder* and *displacive*, implying the former to proceed by a "disordering", and the latter by "displacement", in both cases being a *cooperative* (homogeneous in the bulk) process. However, there was an important footnote in [5]: "There is claim in the literature about connection of emerging rotating molecules (or radicals) in a crystal to second-order phase transitions. That view is erroneous…". Presently the *order-disorder* phase transitions are usually assigned *first* order basically due to a noticeable density "jump", but without realization that they materialize by *nucleation and growth*. The actual crystal rearrangement in an "order-disorder" phase transition is demonstrated in Fig. 3 [11,20]. The details can be found in [20] (section 2.7).

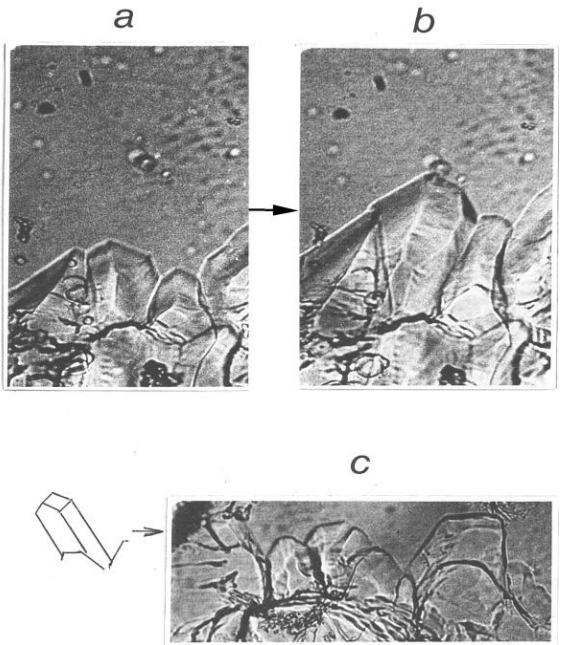

**Figure 5.** Rotational order-disorder L → H phase transition in $CBr_4$. (a,b) Growth of a conglomeration of single crystals (two successive stages). The growing ODCs are not well shaped, but the natural facing is evident. Note that the phase transition is not *cooperative*. The rotational phase (below the interface) and the non-rotational phase (above the interface) merely coexist while all phase rearrangement occurs at the interfaces. It is not a "disordering" in the bulk. (c) Another conglomeration of growing ODCs. Note the ODC reproduced in drawing

## 9. Soft-Mode Phase Transitions

The soft-mode concept was put forward in about 1960 to explain the mechanism of *displacive* ferroelectric transitions, then applied to *order-disorder* ferroelectric transitions and, finally, tried to apply to all "structural" phase transitions. According to the developed theory, a structural phase transition is a cooperative *distortion* of the initial crystal structure as a result of atomic shifts (displacements). This distortion is produced by one of the "soft" (i.e., low-frequency) optical modes, which "softens" toward the transition temperature. When the soft-mode wavelength becomes comparable with the crystal parameters, the cooperative displacement of certain atoms makes the crystal unstable, the displacement suddenly becomes "frozen" and the crystal switches into the alternative phase. The soft-mode concept was developed, tested and demonstrated by using ferroelectric $BaTiO_3$ as an example; even "jumps" in the physical properties at the Curie point were calculated [44]. The same $BaTiO_3$ was used by Landau to illustrate a *continuous* second-order phase transition. Evidently, at least one of these conflicting approaches must be incorrect. But we will set this aside and concentrate on the soft mode model. A first-order phase transition in $BaTiO_3$ is now well established, including all the features of that phase transition type, including large hysteresis of the transition temperature.

In 1970's, the *soft-mode* theory became quite popular [44-50]). Optical and neutron spectroscopic experiments were aimed at finding a soft mode in every phase transition. In 1973 Shirane [50] distinguished two groups of phase transitions in solids: (1) magnetic and superconducting, which he regarded not being "structural" (but they are [20,22]) and maintained that they "were already reasonably understood" (but they were not at that time), and (2) "a large variety of other phase transitions", such as in $SiO_2$, $Nb_2Sn$ and those in ferroelectrics and antiferroelectrics. He contended that "the generalized soft mode concept covers the essential mechanism of phase transitions in solids" and that "the soft mode concept brings a unified picture" of how phase transitions take place in the whole second group known as *structural* phase transitions.

Not only such generalization was premature, the concept itself was not realistic.

1. It fails to comply with the minimal requirements ('A' and 'B' in Section 1) imposed by thermodynamics, considering that its instant *finite* structural "jump" at critical Curie point can only be *infinitesimal*.

2. The *instant* structural "jumps" assumed by the soft-mode concept incorrectly described the real structural phase transitions. The notion that they are instant is possibly rooted in the way Landau used the word "jumps" in describing first-order phase transitions. The irony is that the actual molecule-by-molecule rearrangement is always rather continuous. The "jump" is simply a difference in the structure and properties of the phases coexisting over a temperature range. It looks as a "jump" in the experimental measurements when the temperature range is passed quickly.

3. Being considered second-order, the soft-mode concept should not be applied even to ferroelectric

phase transitions, since "only very few ferroelectrics... have critical or near critical transitions... the majority having first-order transitions" [26]; "most ferroelectric phase transitions are not of second order but first [51]". It remains to add that *all* ferroelectric phase transitions are first order and occur by nucleation and growth.

Then, how can the evidence presented in support of the soft-mode mechanism be explained? It was not definitive at all. In some cases rather "soft" modes were indeed found in the corresponding spectra of a phase, but in many other cases, including almost all molecular crystals [52], no soft modes were detected. Selection of a soft mode that "softens" toward the transition temperature was arbitrary and regarded sufficient to declare the phase transition of the soft-mode type. "Soft" modes, as any vibration modes, can be found in many crystals with or without phase transitions. Like all crystal properties, a soft mode is temperature-dependent and occasionally can show "softening" in the "desirable" direction. This in no way proves that it has any part in the phase transition, if there is one.

The soft-mode concept has not justified the hopes of its inventors. It still exists as one of the possible approaches to some solid-state phase transitions. A truly unified picture of how *all* solid-state phase transitions materialize was described in Section 2..

## 10. Incommensurate Phase Transitions

It had been well established that condensed matter can be in a liquid, crystalline, mesomorphic (liquid-crystalline or orientation-disordered-crystalline) state, or be amorphous. Then the new solid state, called *incommensurate*, was introduced and for a decade or so became very popular in certain circles of research scientists [53-57].

This new solid state was not the subject of interest *per se*. It was invented as a remedy to cure the ailing *soft-mode* model of solid-state phase transitions. Pynn [55] asserted in the 1979 review that "the discovery and study of incommensurably distorted structures is a milestone in the investigation of structural phase transitions". In spite of the word "discovery", no evidence of the *incommensurate* state was presented in that review. As a matter of fact, no hard evidence has ever been found. Yet, the *incommensurate phase transitions* and *incommensurate* solid state were accepted as a reality.

According to the initial *soft-mode* model, a phase transition occurs under the action of a soft mode whose frequency "softens" toward the transition temperature where it turns into zero. There was a problem, however: in most real cases such an optical mode was not found. This increased doubts in the validity of the soft-mode mechanism or, at least, limited its applicability. The new idea was to "soften" requirements to the soft-mode lattice modulation. Now it did not have to "soften" further or even be a rational multiple of a dimension of the crystal unit cell. Now "the new phase does not at all possess any periodicity along the coordinate axis ...; it is referred to as incommensurate. Incommensurability may, naturally, occur along two or three coordinate axes... The fundamental feature of the crystalline state is lost" [55]. The incommensurate phase transition occurs by a "distortion" of the *underlying* ("prototype", "basic", "mother", "undistorted", "symmetrical") higher-temperature phase.

All attention in the literature was directed at the proposed new mechanism of phase transitions. No attention was paid to the resultant peculiar solid state where the displacement of every particular atom had to be unique, so that the resultant structure lacked translation symmetry. Such a solid state defies logic, our knowledge about solid state, and thermodynamics. It cannot exist for any of the following reasons.

(1) The fundamental assumption that structural phase transitions occur by a displacement (distortion, shift) is erroneous, for they occur by nucleation and growth. The relation of the soft-mode and incommensurate transitions to the first/second order classification deserved more attention than a common statement to which class one or another transition belongs. Being a *cooperative* phenomenon, they are usually regarded second-order phase transitions, but applied to first-order and "partly first-order" as well. A first-order incommensurate phase transition is an oxymoron and will not be discussed further. It cannot be of second order either: like the soft-mode transitions it should occur by a *finite* structural jump between the polymorphs and would comply neither with the second-order transitions, which are continuous, nor with thermodynamics.

(2) The theory of a *commensurate* → *incommensurate* transition assumes that the modulating wave becomes "frozen-in" in the resultant phase. The reverse transition could "unfreeze" it, but only with exactly the same mode. However, the vibration spectrum of the resultant phase is different and does not have that particular mode any more. Thus, the conclusion has to be drawn that this type of transition is intrinsically irreversible. What about reversible ones? The theory was silent.

(3) The polymorphs in first-order phase transitions are structurally independent, even according to Landau. But the incommensurate phase transitions assume all the lower-temperature phases of a substance to be derivatives of a "prototype" phase. Suppose there is a prototype high-temperature phase H which changes by a distortion into the lower-symmetric lower-temperature *incommensurate* phase L. The same phase L can also be obtained by growing it from a solution or vapor phase at the lower temperature where it is stable. Then we come to the absurd results: (a) the grown L crystal will have "incommensurate" rather than normal crystal structure, and (b) the grown crystal L will be a *modulated* H phase. Why does the L structure have to be "incommensurate" if the way it came into being had nothing to do with distortion of the "prototype" phase by a vibration mode? What is the source of the "intellect" that enables the crystal grown from solution to know that it must be a distorted version of another phase that can exist at a higher temperature?

(4) The alleged "incommensurate" structure cannot materialize due to a violation of the *close packing principle* valid towards metallic, ionic and molecular crystals. Violation of this principle is equivalent to rejection of the universal principle of minimum free energy in the formation of a structure. Molecular crystals are especially pictorial to illustrate the principle of close molecular packing [58]. The cause behind the principle is minimization of energy of the Van der Waals' interactions in a crystal. By encircling the molecular "skeleton" with the standard Van der Waals' radii, an organic molecule can be assigned a particular shape, as shown in Fig. 6a for biphenyl. Any real organic crystal belongs to one of the most closely packed structures of the molecules defined in this way. For an illustration, the molecular packing of the high-temperature phase of thiourea is shown in Fig. 6b.

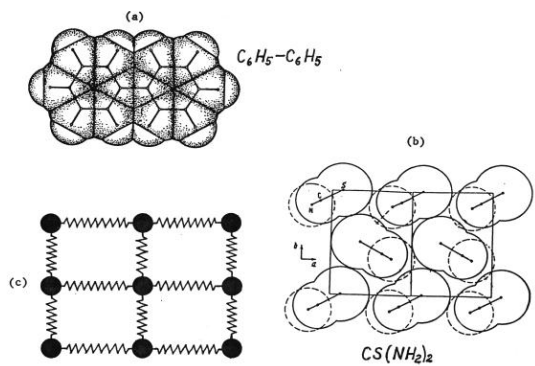

**Figure 6** (a) The model of a biphenyl molecule constructed by encircling the molecular "skeleton" with the intermolecular radii (Kitaigorodskii [58]). (b) The close molecular packing in the high-temperature phase of thiourea. Nitrogen atoms (broken lines) are off the plane *ab* shown by solid lines. The two shown inner molecules have eight "contacts" with the surrounding neighbors (i.e. positioned at the optimum Van der Waals' distances). (c) Any irregular displacements of the balls in this model (equivalent to disturbing the network of standard interatomic distances by an "incommensurate" soft mode in an atomic crystal) will result in returning it into the shown original state. Only rearrangement leading to a new network of standard distances is plausible.

Crystals that disobey the principle of close packing in the "incommensurate" manner are unknown. Incommensurate modulation of a prototype structure by a soft mode will cause individual molecular displacements without regard for the resultant intermolecular distances. Molecules in this structure would penetrate into one another, leaving the adjacent areas vacant. All accumulated experience to date shows that such a structure cannot exist; the polymorphs always represent two different versions of the most closely packed molecules.

To illustrate the point farther, let us turn to the mechanical model of an atomic crystal where balls represent atoms, and springs their bonds (Fig. 6c). To assume that it is possible to produce an "incommensurate" structure from this undistorted structure is equivalent to the assumption that one can displace the balls in different directions (that is, arbitrarily change the lengths of interatomic couplings in the crystal lattice) and the balls will not return to their initial equilibrium positions (i.e., the distortions will be "frozen-in", as a proponent of the incommensurate phase transition would say).

Any particular "incommensurate distortion" depends on the wavelength of the mode that caused this transition ("frozen-in wave"). However, no specific mechanism of phase transition can *impose* the resultant state, because it is determined by the minimal free energy. Its position at the $p-T$ phase diagram is the exclusive function of these parameters, and not the way it arrived there. If the diagram shows the existence of two different crystal phases, the only function of the phase transition, whatever its mechanism is, is to change the above phases from one to other.

Our assertion of the "incommensurate" matter not to exist relates only to the product of the above fictitious phase transition. It does not apply to materials just because someone calls them "incommensurate", for example when some X-ray reflections are found incompatible with the lattice parameters. They resulted from the specific conditions of crystal growth, not phase transition. Thus, a phenomenon comes to mind of a "rythmical" crystal growth from liquid phase, caused by accumulation of latent heat. Another example is "long periods" produced by folding of long-chain molecules. Such imperfect crystal structures do not violate physics of solid-state.

## 11. Scaling Mechanism of Phase Transitions

The modern theoretical physicists in the area of phase transitions pay little attention to the real solid-state phase transitions which materialize by nucleation and crystal growth over a temperature range and exhibit hysteresis. These scientists have their own theoretical world where phase transitions are continuous / homogeneous / critical phenomenon with a fixed ("critical") point to occur and, most importantly, a subject of statistical mechanics.

Such was the "scaling renormalization group" theory of the 1970's, the subject of a Nobel Prize to K. Wilson [59,60]. Even though it was a *theory of second-order phase transitions*, this limitation soon vanished in the same way as it happened to the Landau's theory: it became simply a *theory of phase transitions* [61]. In the instances when first-order phase transitions were not ignored, they were incorporated into the new theory. As one author claimed, "the scaling theory of critical phenomena has been successfully extended for classical first order transitions…" [62]. There is no need to go into the essence of the theory in question. Whether the *scaling* theory could be fruitful in other scientific areas, it has no relation to solid-state phase transitions.

## 12. Quantum Phase Transitions

Specific "quantum" phase transitions were not the product of experimental discovery. They resulted from a theoretical idea. In order to verify legitimacy of their introduction, we turn to the review article "Quantum Phase Transitions" by M. Vojta [63]. His article is helpful on two reasons. (1) It is very authoritative, for S. Sachdev, who had published the canonical book on quantum phase transitions [64], "contributed enormously to the writing of this [Vojta's] article", and many other authorities also had "illuminating conversations and collaborations". (2) The reasons for adding the new class of phase transitions were presented in detail, which made it easier to check them for validity. Several excerptions from the Vojta's article will be used.

Excerpt: *The [non-quantum] phase transitions … occur at finite temperature; here macroscopic order … is destroyed by thermal fluctuations.*

That description of solid state phase transitions is imaginary. It fits to the theory of *continuous* (second-order) phase transitions, but they were not actually found and probably cannot exist at all (see Section 2). Real phase transitions are an intrinsically *local* "molecule-by-molecule" process with the bulks of the coexisting phases remaining static.

Excerpt: [Quantum phase transitions take] *place at zero temperature. A non-thermal control parameter such as pressure, magnetic field, or chemical composition, is varied to access the transition point. There, order is destroyed solely by quantum fluctuations.*

In other words, quantum phase transitions are a version of second-order phase transitions. Replacement of the thermal fluctuations by quantum is considered essential in the theory of quantum phase transitions, but leave the phenomenon to remain "continuous" and occur at "critical points". Now let us place a *real* phase transition near 0°K. The currently relocating molecule (Fig. 1) find itself in the competing attractive fields of forces emanating from the two sides of the interface. The attraction from the side of a lower free energy is stronger. Molecular vibrations, whatever they are, assist in the process, but replacement of thermal fluctuations by quantum fluctuations does not change it. The nucleation and growth will not become the subject of the quantum phase transition theory.

Excerpt: [Classical] *phase transitions are traditionally classified into first-order and continuous transitions. At first-order transitions the two phases co-exist at the transition temperature – examples are ice and water at 0 C, or water and steam at 100 C.*

To the number of different classifications of solid-state phase transitions, the "classical – quantum" was added. How "quantum" phase transitions differ from "classical"? It is not accidental that the chosen examples of first-order phase transitions were not solid-to-solid, even though "quantum" phase transitions are. The reason becomes evident since all "classical" solid-state transitions were assumed "continuous" and a "critical phenomenon". It had to be known that it is not so. It was in direct disregard of L. Landau, who is the author of the "continuous phase transitions" theory: *"Transition between different crystal modifications occurs usually by phase transition at which jump-like rearrangement of crystal lattice takes place and state of the matter changes abruptly. Along with such jump-like transitions, however, another type of transitions may also exist…"* [5]. Thus, phase transitions between crystal modifications are *first order*, but "continuous" phase transitions only *may* exist. As noted in the introduction, sufficiently documented second-order phase transitions were not found. The two phases in the *real* "classical" solid-state phase transitions coexist over a temperature range, and not only at a single temperature point.

The theory of quantum phase transitions calls all solid-state phase transitions away from 0°K "classical". Even though they are not named "second-order" in the Vojta article on some unexplained reason, they are deemed "continues" and occur at their critical points where the previously existing order is destroyed by thermal fluctuations. Toward 0°K the thermal fluctuations fade away, while the quantum fluctuations take over. The "classical" critical points become "quantum" critical points. The conclusion about existence of the "quantum" brand of phase transitions are ruined as soon as it is clarified that the "classical" phase transitions are a *nucleation and growth*. There are no critical points. The premise was erroneous.

Even though the point is now proven, it is useful to extend the analysis somewhat further.

Excerpt*: In contrast, at continuous transitions the two phases do not co-exist. An important example is the ferromagnetic transition of iron at 770 C, above which the magnetic moment vanishes. This phase transition occurs at a* *point where thermal fluctuations destroy the regular ordering of magnetic moments – this happens continuously in the sense that the magnetization vanishes continuously when approaching the transition from below. The transition point of a continuous phase transition is also called critical point.*

Ferromagnetic phase transitions had become the last resort for the conventional theory to exemplify "continuous" phase transitions and critical phenomena. The above contradictory explanation (magnetization changes continuously at critical point) illustrates the problem to treat them as second order. It has been shown [20] (Chapter 4), [22] that they too materialize by crystal growth. As for ferromagnetic transition of *Fe*, a "discontinuity" of the Mössbauer effect there was reported already in 1962 by Preston [65,66], who stated that this "might be interpreted as evidence for a first-order transition". It was analyzed in [20] (Sec. 4.2.3, 4.7) and concluded to be a case of nucleation and growth. Finally, the first order ferromagnetic phase transitions in *Fe, Ni* and *Co* were confirmed by recording their latent heat [67].

To complete the picture, there were publications where certain "quantum" phase transitions were stated to be first order. Evidently, some authors must be incorrect. Who it was: those arguing the "quantum" phase transitions to be a "critical phenomenon" and the antithesis to first-order phase transitions, or those embracing "first-order quantum phase transitions"? The answer is: all of them are. The experimentalists, who concluded their "quantum" phase transitions being first order, are less erroneous. Their "quantum" phase transitions were first-order indeed, just not being "quantum".

## 13. Conclusion

Solid-state phase transitions were a mystery over almost all 20[th] century, extended to the 21[st] for those who do not know about the already found solutions. All that time was marked by a succession of the theories, all based on the "cooperative" idea, each one after disappointment in the previous theory. But neither theory is being completely abandoned, while the "quantum" phase transitions is still rather popular.

It is understandable how exciting it was for experimentalists to discover such anomalies as the λ-peaks, for they seemed to promise a breakthrough in a previously unexpected direction. It was not less exciting for theoretical physicists to find in the anomalies the area of application of their talents, knowledge of statistical mechanics and belief in its general power and dynamical nature of everything. But Nature had its own agenda, namely, to make its natural processes (a) universal, (b) simple and (c) the most energy-efficient. Being uncompromising in these principles, Nature produced better processes than most brilliant human beings, even Nobel Prize Laureates, could invent.

Solid-state phase transition is such a process. It is more universal, simple and energy-efficient than statistical-dynamic theories could offer. It is universal because it is just a particular manifestation of the general crystal growth. It is also as simple as crystal growth. It is energy-efficient because it needs energy to relocate one molecule at a time, and not the myriads of molecules at a time as a cooperative process requires.

An important lesson can be drawn from this. The whole effort was largely misdirected. Great amounts of time, hard work, resources and talent were wasted. Insufficient attention to facts, such as the disregard of the nucleation and growth as a mechanism inherent in all solid-state phase transitions, was substituted by excessive theoretical creativity. The contradictions were tolerated, while correct solutions were ignored. "Tries and errors" is a normal way of a scientific advancement; it is only honorable to recognize being incorrect. But that has not happened (yet?) in the area of solid-state phase transitions. As a result, the general understanding of how they materialize was unnecessarily delayed for very long time.

______________________________________________